\documentclass{sig-alternate}
\usepackage{hyperref}

\begin{document}

\conferenceinfo{WIMS'11, May 25-27, 2011}{Sogndal, Norway}
\CopyrightYear{2011} 
\crdata{978-1-4503-0148-0/11/05}  

\title{Crawling Facebook for Social Network Analysis Purposes}

\numberofauthors{5} 
\author{
\alignauthor
Salvatore A. Catanese\\
	\affaddr{Dept. of Physics\\ Informatics Section}\\
	\affaddr{University of Messina, Italy}\\
	\email{salvocatanese@gmail.com}
\alignauthor
Pasquale De Meo\\
    \affaddr{Dept. of Physics\\ Informatics Section}\\
    \affaddr{University of Messina, Italy}\\
    \email{demeo@unirc.it}	
\alignauthor
Emilio Ferrara\\
    \affaddr{Dept. of Mathematics}\\
    \affaddr{University of Messina, Italy}\\
    \email{emilio.ferrara@unime.it}
\and 
\alignauthor
Giacomo Fiumara\\
    \affaddr{Dept. of Physics\\ Informatics Section}\\
    \affaddr{University of Messina, Italy}\\
    \email{giacomo.fiumara@unime.it}
\alignauthor
Alessandro Provetti\titlenote{A. P. is also long-term visitor to the Oxford-Man Institute, University of Oxford, UK.}\\
  \affaddr{Dept. of Physics\\ Informatics Section}\\
 	\affaddr{University of Messina, Italy}\\
	\email{ale@unime.it}       
}

\date{30 September 2010}
\maketitle

\begin{abstract}
We describe our work in the collection and analysis of massive data describing the connections between participants to online social networks. 
Alternative approaches to social network data collection are defined and evaluated in practice, against the popular Facebook Web site.
Thanks to our ad-hoc, privacy-compliant crawlers, two large samples, comprising millions of connections, have been collected; the data is anonymous and organized as an undirected graph. 
We describe a set of tools  that we developed to analyze specific properties of such social-network graphs, i.e., among others, degree distribution, centrality measures, scaling laws and distribution of friendship.
\end{abstract}

\category{H.2.8}{Database Management}{Database Applications}[Data mining]
\category{E.1}{Data Structures}[Graphs and networks]
\category{G.2.2}{Discrete Mathematics}{Graph Theory}[Network problems]

\terms{Web data, Design, Experimentation, Human Factor}


\section{Introduction}
The increasing popularity of online Social Networks (OSNs) is witnessed by the huge number of users acquired in a  short amount of time: some social networking services now have gathered hundred of millions of users, e.g. Facebook, MySpace, Twitter, etc. 
The growing accessibility of the Internet, through several media, gives to most of the users a 24/7 online presence and encourages them to build a solid online interconnection of relationships. 
As OSNs become the tools of choice for connecting people, sociologists expect that their structure will increasingly mirror real-life society and relationships. 
At the same time, with an extimate 13 millions transactions per seconds (at peak) Facebook is one of the most challenging computer science artifacts, posing several optimization and robustness challenges. 
The analusis of OSN connection is of scientificic interest on multiple levels. 
First of all, large scale studies of models reflecting a real community structure and its behavior were impossible before. 
Second, data is clearly defined by some structural constraints, usually provided by the OSN structure itself, with respect to real-life relations, often hardly identifiable. 

Computer science covers a fundamental role in this perspective, as it provides tools and techniques to automatize the process of gathering data from massive data sources and their analysis, at several levels of granularity. 
To analyse such a complex object, proper metrics need to be introduced, in order to identify and evaluate properties of the considered social network. 
We will highlight the role of the SNA as an indispensable step to achieve a complete knowledge in this area. 

Identifying drawbacks is important in order to acquire a correct methodology. 
Because of the complexity of this kind of networks, investigating the scalability of the problem is crucial: we do not have computational resources able to mine and work with the whole Facebook graph, for several reasons. 
First of all, it is not trivial to tackle large scale mining issues: for example, last year Gjoka et al. \cite{Gjoka2010} measured the crawling overhead in order to collect the whole Facebook graph in 44 Terabytes of data to be downloaded and handled. 

Moreover, even when such data can be be acquired and stored locally, it is non-trivial to devise and implement functions that traverse and visit the graph or even evaluating simple metrics.
For all these reasons it is common to work with small but representative sample of the graph. 
In literature, extensive research has been conducted on sampling techniques for large graphs but, only in the last few years, some studies have sehd light on the partial bias that standard methodologies introduce, one way or another.

This paper is organized as follows: Section 2 presents the summary of the most representative related work.
Section 3 describes the methodology we used to conduct this work, in particular identifying some points of interest to be inspected, establishing algorithms and techniques to be exploited and defining goals and achievements. 
Section 4 focuses on the data collection process and related issues: we define the technical challenges underlying the process of information extraction from Facebook, and describe in detail the design and implementation of our application, called \textit{crawling agent.}
Once it is gathered, OSN data needs to be analyzed and several experimentation details regarding SNA aspects are discussed in Section 5. 
The most important results are summarized in Section 6, where we introduce some measures obtained when SNA metrics is applied  to our dataset. 
Finally, Section 7 summarizes our work and discusses a blueprint for future work on Facebook analysis.

\section{Related Work}
The classic work on social networks is rooted in the field of \emph{Sociometry}; in late sixties Milgram and Travers \cite{Travers1969} introduced the well-known theories of the \emph{six-degrees of separation} and the \emph{small world}.
Zachary \cite{Zachary1980}, in his PhD thesis, formalized the first model of a real-life social network. 
Then, social networks started attracting the interest of different sciences, including Computer Science; for example Kleinberg \cite{Kleinberg2000} studied the algorithmic aspects of social networks while Staab et al. \cite{Staab2005} analyzed their range of applicability.

Nowadays, also thanks to the visibility of OSNs, the scientific research in this field counts several \textit{hot topics}.
It is possible to identify at least three distinct, not necessarily disjoint, directions of research; they focus, respectively, on i) data collection techniques, ii) characterization of online social networks and iii) online social networks analysis.

First, Leskovec and Faloutsos \cite{Leskovec2006} cover sampling techniques applied to graphs, exploiting several algorithms for efficiently visiting large graphs and avoiding bias of data. 
Several work covers aspects regarding crawling techniques for mining social networks: Ye et al. \cite{Ye2010} provide an exhaustive walk-through about this field. 

With respect to data collection aspects, the work by Gjoka et al. \cite{Gjoka2010} on OSNs (in particular on Facebook) is closely related to our work.

In this paper we describe in details several results, some of them already presented in a preliminary work \cite{Catanese2010} based on a significantly smaller sample acquired with a naive technique.  

Chau et al. \cite{Chau2007} focus on the parallel crawling of OSNs; Gjoka et al. \cite{Gjoka2010} refer they used a parallel crawler written in Python and running on a cluster of 56 machines, but avoiding technical details.

In the second category we find works whose main goal is to discover properties of online social networks. 
Sometimes companies provide the complete OSN dataset, e.g. Ahn et al. \cite{Ahn2007} studied the graph of a South Korean OSN, named CyWorld, and its scalability properties, in order to correctly estimate degree distribution and clustering coefficient in smaller samples. 
Leskovec \cite{Leskovec2008}, in his PhD thesis, studied dynamics of large social networks analyzing data provided by Microsoft about the usage of their instant messaging platform, formerly called MSN. 
More frequently, social network services companies like Facebook are reluctant to share their data for research purposes. 
The only viable solution is to acquire this information crawling the front-end of the social networks, wrapping and managing public accessible data. 
Ferrara et al. \cite{Baumgartner2010} provided an exhaustive survey of these approaches, commonly referred to as Web Information Extraction techniques. 
Wilson et al. \cite{Wilson2009} defined a \emph{region-constrained} breadth-first-search sampling methodology applied to the Facebook social graph and formalized some properties as \textit{assortativity} and \textit{interaction}, easily verifiable in small regions, but not generalizable to the whole graph.

Work belonging to the last category usually present SNA techniques applied to OSNs. 
Carrington et al. \cite{Carrington2005} formalized a rigorous methodology to model OSNs, in particular to discover, if existing, aggregations of nodes covering particular positions in the graph (e.g. centrality positions) and clusters of particular relevance. 
Kumar et al. \cite{Kumar2009,kumar2010structure} focused on analyzing the structure of the \textit{giant component} of the graph, trying to define a generative model to describe the evolution of the network; they also introduced techniques of comparison of simulations against actual data to verify the reliability of the model. 
Mislove et al. \cite{Mislove2007} illustrated several aspects of the analysis of OSNs; they crawled large OSNs, then analyzed several networks using methods formalized in the SNA, e.g. inspecting link symmetry, power-law node degrees, groups formation, etc. 

Finally, literature on analysis of the behavior of OSNs users is abundant: Golbeck and Hendler \cite{Golbeck2006} described the evolution of trust relationships among them.
Liben-nowell and Kleinberg \cite{Liben-nowell2007} studied the link-prediction problem, formalizing the concept of proximity of networks users. 
Several studies characterized OSNs users behaviors \cite{Gjoka2008,Maia2008,Benevenuto2009}, thus Schneider et al. \cite{Schneider2009} provided a starting point illustrating how to analyze OSNs (in particular Facebook) from a network perspective.

\section{Methodology}
OSNs can be represented as graphs, nodes represent users, edges represent connections. 
Facebook, in particular, is characterized by a simple friendship schema, so as it is possible to represent its structure through an unweighted, undirected graph. 

In order to understand if it is possible to study OSNs without having access to their complete dataset (e.g. Facebook), we adopted two different methodologies to gather partial data (i.e. a sub-graph of the Facebook complete graph).

Our purpose is to acquire comparable samples of the Facebook graph when applying different sampling methodologies. 
Once collected, information are analyzed using tools and techniques provided by the SNA. 
We also investigate the scalability of this problem, evaluating several properties of different graphs and sub-graphs, collected with different sampling approaches. 

Then, we compare measures calculated applying SNA techniques with statistical information provided by Facebook itself in order to validate the reliability of collected data. 
Moreover, highlighting similarities and differences between results obtained through this study and similar ones, conducted in past years, is helpful to understand the evolution of the social network over the time.

Finally, this methodology starts from the assumption that, to gain an insight into an OSN, we take a \emph{snapshot} of (part of) its structure. 
Processes of data collection require time, during which its structure could slightly change, but, assuming the small amount of time, with respect to the whole life of the social network, required for gathering information, this evolution can be ignored. Even though, during the data cleaning step we take care of possible information discrepancies.

\subsection{Breadth-first-search sampling}
Breadth-first-search (BFS) is a well-known graph traversal algorithm proved to be optimal and easy to be implemented, in particular for visiting unweighted, undirected graphs. 
For these reasons it has been adopted in several OSNs crawling tasks \cite{Mislove2007,Chau2007,Wilson2009,Gjoka2010,Ye2010,Catanese2010}.

Starting from a \emph{seed} node, the algorithm discovers first neighbors of the seed, putting them in a FIFO queue. 
Nodes in the queue are visited in order of appearance, so the coverage of the graph could be represented as an expanding wavefront. 
This algorithm, virtually, concludes its execution when all discovered nodes have been visited. 
For large graphs, like OSNs, this stopping condition would imply huge computational resources and time. 
In practice, we established as termination criteria, a coverage of at least three sub-levels of friendships and a running time of 240 hours, so as resulting in an incomplete visit of the graph.

Kurant et al. \cite{kurant2010bias} assert that an incomplete BFS sampling leads to biased results, in particular towards high degree nodes. 
Even though in our experimentation data acquired through BFS sampling do not show a statistically significant bias, we investigate this aspect in comparison with others, obtained using a sampling technique which is proved to be unbiased. 

\subsection{Uniform sampling}
The \emph{Uniform} sampling of Facebook has been introduced by Gjoka et al. \cite{Gjoka2010}; they provided proof of correctness of this approach and implementation details, omitted here. 

Briefly, Facebook relies on a well-designed system of user-IDs assignment, spreading in the space of 32-bit range, so as the commonly called \emph{rejection sampling} methodology is viable. 
This approach requires the generation of a queue of random user-IDs to be requested to Facebook, querying for their existence. 
If so, the user and his/her personal friend list are extracted, otherwise the user-ID is discarded and the polling proceeds with the next. Advantages of this approach rely on the independence of the distribution of user-IDs with respect to the distribution of friendships in the graph.

\section{Data Collection}
\begin{figure}[ht]
	\includegraphics[width=\columnwidth]{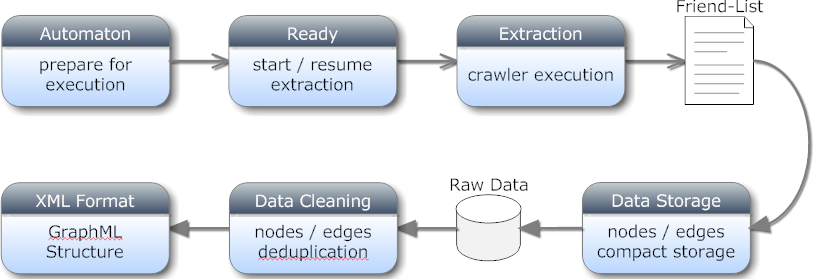}%
	\caption{State diagram of the data mining process}%
	\label{fig1}%
\end{figure}

Regardless of the methodology implemented, our process of data collection can be schematized as follows (see Figure \ref{fig1}):
\begin{enumerate}
	\item Preparation for the execution of the agent.
	\item Starting or resuming the process of data extraction.
	\item The crawler execution extracts friend lists, cyclically.
	\item Raw data are collected until the extraction process concludes or it is stopped.
	\item Data cleaning and de-duplication of information.
	\item Eventually, data structured in GraphML format.
\end{enumerate}

\subsection{Privacy policies}
Companies providing social network services like Facebook, build their fortune on behavioral and targeted advertising, exploiting the huge amount of information they own about users and their activity to show them thematized ads, so as to increase visibility and earnings of advertised companies. 
This is why they are reluctant to share information about users and the usage of the network. 
Moreover, several questions related to users privacy are adducted \cite{Gross2005,McCown2009}. 

For all these reasons, Facebook, like many other common social network services, is implemented as a black box. 
Users can modify their privacy setting and Facebook ensures several levels of information hiding; by default, users can only access private information of other users belonging to their friendship network, while friend list pages are publicly accessible. 

To maintain this \emph{status quo}, Facebook implements several rules, some behavioral (e.g. \emph{terms of use} prohibits data mining, regardless of the usage of data extracted) and some technicals (e.g the friend list is dispatched through a script which asynchronously fills the Web page, preventing naive techniques of data extraction).

\subsection{Sampling sessions of Facebook}

\subsubsection{BFS crawler}
The architecture of this crawler includes an agent that executes data extraction tasks and a FIFO queue, named \emph{ToBeVisited}, in which are stored, in order of appearance, profiles of users to be visited.

The flow of HTTP requests sent by the BFS crawler is described as follows: first, the agent contacts the Facebook server, providing credentials required for the authentication through \emph{cookies}. 
Once logged in, the agent starts crawling pages, visiting the friend list page of the \emph{seed} profile (the logged in user) and extracts her friend list; friends user-IDs are enqueued in a \emph{to be visited} FIFO queue and, cyclically, they are visited in order to retrieve their friend list (see Table \ref{tab1}). 

We started this process from a single seed, and stopped its execution after 10 days of crawling, obtaining a partial sample of the Facebook graph and contains information down to the third sub-level of friendships (friend lists of friends of friends have been acquired). Datasets are described in a following specific section.

\begin{table*}
	\centering
	\begin{tabular}{|c| l c c l c|}
	\hline
	 \textbf{N.} & \textbf{Action} & \textbf{Protocol} & \textbf{Method} & \textbf{URI} & \textbf{KBytes} \\
	\hline
	\hline
  1 & \multicolumn{5}{l|}{open the Facebook page} \\
		& & HTTP & GET & www.facebook.com/ & 242 \\
	2 & \multicolumn{5}{l|}{login providing credentials} \\
		& & HTTPS & POST & login.facebook.com/login.php?login\_attempt=1 & 234 \\
		& & HTTP & GET & /home.php & 87 \\
	3 & \multicolumn{5}{l|}{open seed profile friend list} \\
		& & HTTP & GET & /friends/?filter=afp & - \\
		& & HTTP & GET & /friends/ajax/friends.php?filter=afp & 224 \\
	\hline
	\hline
	4 & \multicolumn{5}{l|}{extract the friend list through regular expressions, put profiles in the queue \emph{to be visited}} \\
	5 & \multicolumn{5}{l|}{visit the next profile in the queue \emph{to be visited}} \\
	  & & HTTP & GET & /friends/?id=\emph{XXX}\&filter=afp & - \\
		& & HTTP & GET & /friends/ajax/friends.php?id=\emph{XXX}\&filter=afp & 224 \\
	6 & \multicolumn{5}{l|}{cycle the process going to step 4} \\
	\hline
	\end{tabular}
	\caption{HTTP requests flow of the BFS crawler for connecting to Facebook and retrieving information.}
	\label{tab1}
\end{table*}

\subsubsection{Uniform crawler}
The Uniform crawler reflects an architecture identical to the BFS crawler (refer to \ref{tab1} for details). 
The only difference is the queue generation and management.

The most important aspect for ensuring the efficiency of this crawler is related to the comparability of all the possible assigned users-IDs with the actual assigned user-IDs.
As of August 2010, when we crawled the graph, Facebook declared more than half a billion users; so that the number of users with a 32-bit user-ID assigned is approximately $2^{29} \simeq 5.37e9$. 
Subsequently, this approach ensures that the crawler finds an existing user profile, statistically every $\frac{2^{32}}{2^{29}} = 2^3 = 8$ attempts, a result even better than the situation reported by \cite{Gjoka2010}. 

Our purpose is to extract a uniform sample whose dimensions are comparable with the BFS sample extracted, which contains 63.4K unique visited users. Statistically, every eight attempts one existing user is matched, so we generated eight queues of $2^{16} \simeq 65.5K$ user-IDs, randomly extracted in the interval $[0, 2^{32}-1]$.
These queues fed eight different agents which crawled Facebook for 10 days. 
The total expected number of users was $2^{16} \simeq 65.5K$.

Any discrepancies between the total number of users acquired and the theoretic number of possible acquirable users is due to the privacy policy setting preventing friend lists to be visited anonymously. 
We investigated also this aspect.

\subsubsection{Limitations}
One notable limitation we met during the data mining process is due to the technical precautionary adopted by Facebook to avoid an high traffic through their platform.
In details, once the agent requests for a friend-list Web page, the server responds with a list of at most 400 friends. 
If the list should contain more than 400 friends, it is shortened to 400 instead.

This limitation can be avoided adopting different techniques of data mining, for example exploiting platforms of Web data extraction which scrape information directly from the Web page, simulating the behavior of a human user, through the interface of the browser itself.
Even though, the computational overhead of a similar process is too high to be adopted for a large-scale crawling task, unless a commercial platform for Web data extraction is employed.

On a smaller scale, we already faced the problem of sampling Facebook adopting this approach \cite{Catanese2010}.

\subsection{Data cleaning}
Data cleaning and post-processing aspects represent an important step in a sampling task. 
During the process of data collection it is possible, in our specific case, that multiple instances of the same edge and/or parallel edges could be stored (this because the Facebook graph is undirected, so one edge connecting two nodes is sufficient to correctly represent the structure of the network). 
This information is redundant and unnecessary.

We adopted hash functions to efficiently remove duplicate edges from our datasets, in a linear time $O(n)$, with respect to the dimension of the datasets.
Moreover, because Facebook allows users to substitute their numerical ID with an alphanumerical one, we obtained datasets containing both these two formats of IDs.

We decided to conform all the user-IDs to a same numerical format, and, in order to avoid collisions with the 32-bit coding system adopted by Facebook, we chose the 48-bit hybrid additive-rotative hashing algorithm by Partow \cite{Partow}, to convert all the user-IDs.
Thus, we consequently obtained also anonymized datasets: user-IDs are ``encrypted'' adopting the 48-bit hashing to obtain data with no references to the users.
Moreover, to comply with the Facebook end-user licence, we never memorize users' sensible data.

In conclusion, we verified the integrity and the consistency of data.
Some particular sub-graphs (namely, ego-networks) of interest, have been converted in the GraphML format \cite{Brandes2002}, for additional experimentation, adopting tools which require this standard for data management.

\subsection{Datasets}
A quantitative description of datasets follows in Table \ref{tab2}.

\begin{table}[h]
	\centering	
	\begin{tabular}{| l | c c |}
		\cline{2-3}
		\multicolumn{1}{r|}{} & \textbf{BFS} & \textbf{Uniform} \\
		\cline{2-3}
		\hline
		Queue length & 11.59M & $8 \cdot 2^{16}$ = 524K\\
		N. of unique visited users & 63.4K & 48.1K \\
		N. of unique disc. neighbors & 8.21M & 7.69M\\
		N. of unique edges & 12.58M &  7.84M\\
		Avg. degree & 396.8 & 326.0\\
		Median degree & 400 & 400 \\
		Largest eigenvector value& 68.93 &23.63\\
		Effective diameter & 8.75& 16.32\\
		Largest component has & 99.98\% & 94.96\%\\
		Avg. clustering coefficient  & 0.0789 & 0.0471 \\
		\hline
		\hline
		Crawling period & 08/01-10 & 08/11-20 \\	
		\hline
		\end{tabular}	
	\caption{Datasets acquired and analyzed.}
	\label{tab2}
\end{table}

\subsubsection{BFS dataset}
For the first sampling task we adopted the BFS algorithm, during the first ten days of August 2010.
The collected sample contained 12.58 millions edges connecting 8.21 millions users (i.e. the 29.2\% of the nodes was duplicates); the complete 2.0-degree ego-networks (i.e. all the complete sub-graphs in which we visited the whole neighborhood of a specific node) were 63.4 thousands, and we adopted this measure as yardstick for the Uniform sampling task.

Some of the most interesting properties of this sample are the following: the average degree of this BFS sample is 396.8, a values conform to the average depicted by similar studies \cite{Gjoka2010,Wilson2009}. 
The median value, as already discussed, is affected by the technical limitation imposed by Facebook.
The effective diameter of this graph is small; although this is completely compatible with theories like ``the small world phenomenon'', we suppose that the measure could be affected by the particular behavior of the incomplete BFS traversal, which radially covers the graph.
The largest connected component includes almost all the nodes, while the clustering coefficient of the network (0.0789) is perfectly alligned to the interval already evaluated by the previously mentioned studies ([0.05 - 0.18]).

\subsubsection{Uniform dataset}
Our purpose for the Uniform sampling was to acquire a dataset whose dimensions may be comparable with the BFS dataset. 

We considered the following criteria to define how to proceed: first of all, we would like to exploit the intrinsic nature of this parallelizable problem, adopting several different crawlers working at the same time.
Following the past considerations that the probability of matching an existing user with this technique is $\frac{1}{8}$, we generated eight different queues, feeding eight crawlers.
Each queue was composed of $2^{16} \simeq 65.5K \cong 63.4K$, our yardstick.

During the second ten days of August 2010, these crawlers collected a sample of 7.69 millions users connected each others through 7.84 millions edges.
The following statistics highlights the quality of the sample: the average degree is smaller with respect to the BFS sample and it probably better reflect the expected value.
Once again the clustering coefficient, although smaller with respect to the BFS sample, is acceptable.

Different considerations hold about the effective diameter and the largest connected component: the first is still too high, but this measure is probably affected by the small nature of the sample. 
This possible explanation is supported by the evidence that the largest connected components do not include about the 5\% of nodes, which remain disconnected (because of the randomness of the sampling process).

\section{Results}
The discussion of experimental results follows.
The analysis of collected datasets has been conducted exploiting the functionalities of the  Stanford Network Analysis Platform library (SNAP) \cite{snap}, which provides general purpose network analysis functions.


\subsection{Overall metrics}
Metrics previously presented in the analysis of datasets are part of the methodology defined, e.g. by Carrington et al. \cite{Carrington2005}, for social network analysis tasks.
Another interesting perspective, for the same purpose, is the list of metrics to be analyzed, provided by Perer and Shneiderman \cite{perer2006balancing}.

The following measures have been investigated.

\subsubsection{Degree distribution}
The first interesting  property we analyzed is the degree distribution, which is reflected by the topology of the network.
The literature refers that social networks are usually described by power-law degree distributions, $P(k)\sim k^{-\gamma}$, where $k$ is the node degree and $\gamma \le 3$.
Networks which reflect this law share a common structure with a relatively small amount of nodes connected through a big number of relationships.

The degree distribution can be represented through several distribution function, one of the most commonly used being the Complementary Cumulative Distribution Function (CCDF), $\displaystyle{\wp (k) = \int_k^\infty P(k')dk' \sim k^{-\alpha} \sim k^{-(\gamma -1)}}$, as reported by Leskovec \cite{Leskovec2008}.

\begin{figure}[!ht]%
	\includegraphics[width=\columnwidth]{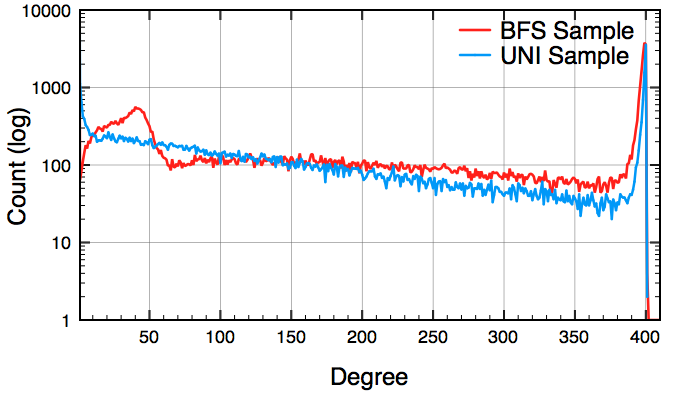}%
	\caption{BFS vs. UNI: degree distribution}%
	\label{outDeg}%
\end{figure}

In Figure \ref{outDeg} we show the degree distribution based on the BFS and Uniform (UNI) sampling techniques.
The limitations due to the dimensions of the cache which contains the friend-lists, upper bounded to 400, are evident.
The BFS sample introduces an overestimate of the degree distribution in the left and the right part of the curves.

\begin{figure}[!ht]%
	\includegraphics[width=\columnwidth]{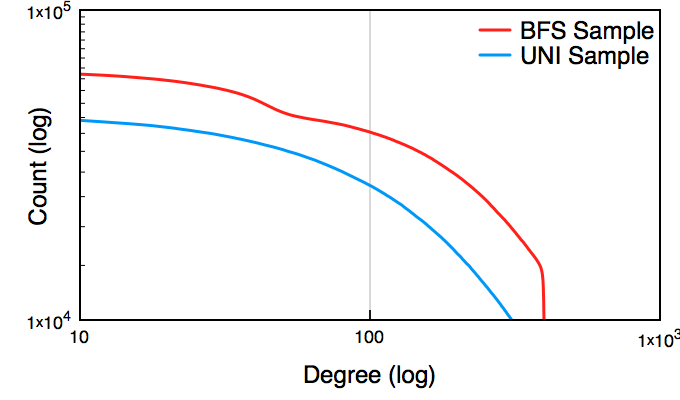}%
	\caption{BFS vs. UNI: CCDF degree distribution}%
	\label{outDegCCDF}%
\end{figure}

The Complementary Cumulative Distribution Function (CCDF) is shown in Figure \ref{outDegCCDF}.

\subsubsection{Diameter and hops}
Most real-world graphs exhibit relatively small diameter (the ``small-world'' phenomenon \cite{milgram1967small}, or ``six degrees of separation''): a graph has diameter D if every pair of nodes can be connected by a path of length of at most D edges.
The diameter D is susceptible to outliers.
Thus, a more robust measure of the pairwise distances between nodes in a graph is the effective diameter, which is the minimum number of links (steps/hops) in which some fraction (or quantile q, say q = 0.9) of all connected pairs of nodes can reach each other.
The effective diameter has been found to be small for large real-world graphs, like Internet and the Web \cite{albert1999diameter}, real-life and Online Social Networks \cite{albert2002statistical,leskovec2005graphs}.

Hop-plot extends the notion of diameter by plotting the number of reachable pairs $g(h)$ within $h$ hops, as a function of the number of hops \emph{h} \cite{palmer2002generating}.
It gives us a sense of how quickly nodes' neighborhoods expand with the number of hops.

\begin{figure}[!ht]%
	\includegraphics[width=\columnwidth]{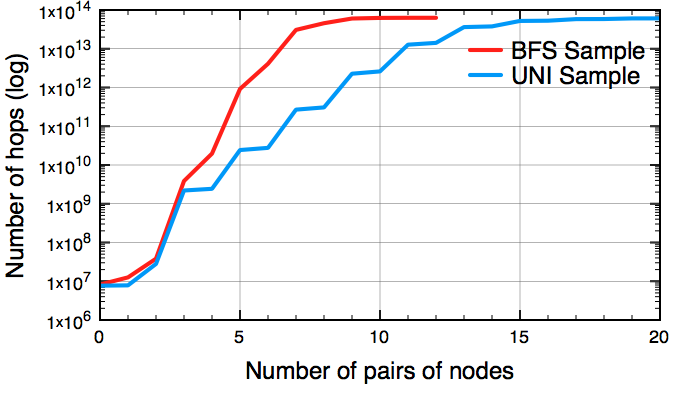}%
	\caption{BFS vs UNI: hops and diameter}%
	\label{hop}%
\end{figure}

In Figure \ref{hop} the number of hops necessary to connect any pair of nodes is plotted as a function of the number of pairs of nodes.
As a consequence of the more ``compact" structure of the graph, the BFS sample shows a faster convergence to the asymptotic value listed in Table \ref{tab2}.

\subsubsection{Clustering Coefficient}
The clustering coefficient of a node is the ratio of the number of existing links over the number of possible links between its neighbors.
Given a network $G = (V,E)$, a clustering coefficient, $C_i$, of node $i \in V$ is:
\[
C_i = 2|\{(v, w)|(i, v), (i, w), (v, w) \in E\}|/k_i(k_i-1)
\]
where $k_i$ is the degree of node $i$. 
It can be interpreted as the probability that any two randomly chosen nodes that share a common neighbor have a link between them.
For any node in a tightly-connected mesh network, the clustering coefficient is 1.
The clustering coefficient of a node represents how well connected its neighbors are.

Moreover, the overall clustering coefficient of a network is the mean value of clustering coefficient of all nodes.
It is often interesting to examine not only the mean clustering coefficient, but its distribution.

\begin{figure}[!ht]%
	\includegraphics[width=\columnwidth]{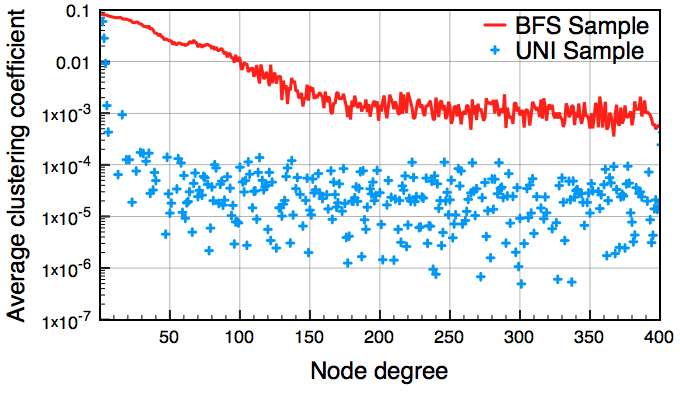}%
	\caption{BFS vs. UNI: clustering coefficient}%
	\label{ccf}%
\end{figure}

In Figure \ref{ccf} is shown the average clustering coefficient plotted as a function of the node degree for the two sampling techniques.
Due to the more "systematic" approach, the BFS sample shows less violent fluctuations.

\subsubsection{Futher Considerations}
The following considerations hold for the diameter and hops: data reflected by the BFS sample could be affected by the ``wavefront expansion'' behavior of the visiting algorithm, while the UNI sample could still be too small to represent the correct measure of the diameter (this hypothesis is supported by the dimension of the largest connected component which does not cover the whole graph, as discussed in the next paragraph).
Different conclusions can be discussed for the clustering coefficient property.
The average values of the two samples fluctuate in the same interval reported by similar studies on Facebook (i.e. [0.05, 0.18] by \cite{Wilson2009}, [0.05, 0.35] by \cite{Gjoka2010}), confirming that this property is preserved by both the adopted sampling techniques.

\subsubsection{Connected component}
A connected component or just a component is a maximal set of nodes where for every pair of the nodes in the set there exist a path connecting them.
Analogously, for directed graphs we have weakly and strongly connected components.

\begin{figure}[!ht]%
	\includegraphics[width=\columnwidth]{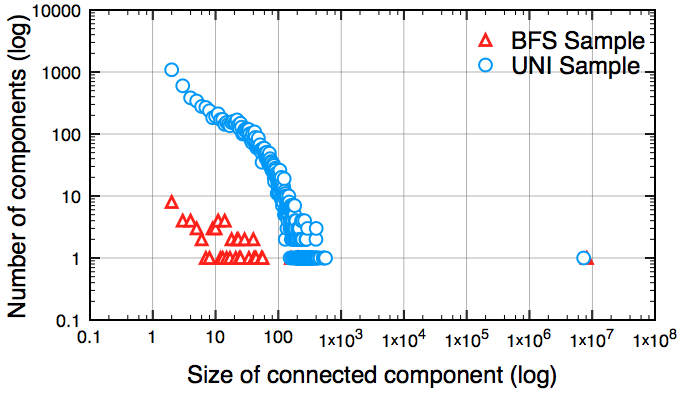}%
	\caption{BFS vs. UNI: connected component}%
	\label{wcc}%
\end{figure}

As shown in Table \ref{tab2}, the largest connected components cover the 99.98\% of the BFS graph and the 94.96\% of the UNI graph.
This reflects in Figure \ref{wcc}.
The scattered points in the left part of the plot have a different meaning for the two sampling techniques.
In the UNI case, the sampling picked up disconnected nodes.
In the BFS case disconnected nodes are meaningless, so they are probably due to some collisions of the hashing function during the de-duplication phase of the data cleaning step.
This motivation is supported by their small number (29 collisions over 12.58 millions of hashed edges) involving only the 0.02\% of the total edges.
However, the quality of the sample is not affected.

\subsubsection{Eigenvector}
The eigenvector centrality is a more sophisticated view of centrality: a person with few connections could have a very high eigenvector centrality if those few connections were themselves very well connected.
The eigenector centrality allows for connections to have a variable value, so that connecting to some vertices has more benefit than connecting to others.
The Pagerank algorithm used by Google search engine is a variant of eigenvector Centrality.

\begin{figure}[!ht]%
	\includegraphics[width=\columnwidth]{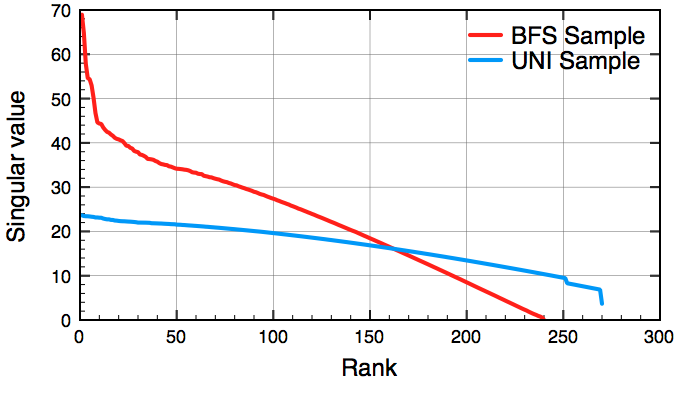}%
	\caption{BFS vs. UNI: singular values}%
	\label{sngVal}%
\end{figure}

\begin{figure}[!ht]%
	\includegraphics[width=\columnwidth]{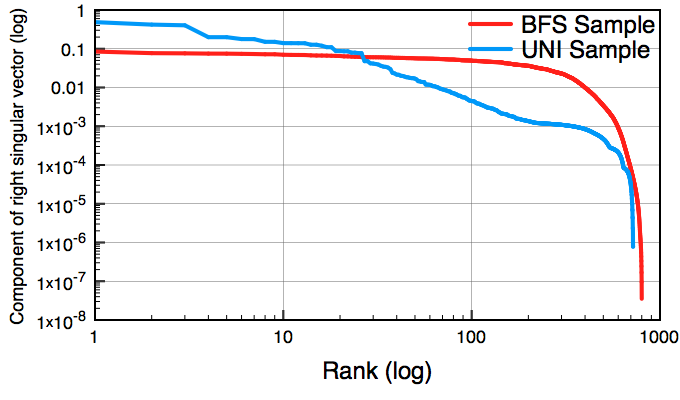}%
	\caption{BFS vs. UNI: right singular vector}%
	\label{sngVecL}%
\end{figure}

Figure \ref{sngVal} shows the eigenvalues (singular values) of the two graphs adjancency matrices as a function of their rank.
In Figure \ref{sngVecL} is plotted the Right Singular Vector of the two graphs adjacency matrices as a function of their rank using the logarithmic scale.
The curves relative to the BFS sample show a more regular behavior, probably a consequence of the richness of the sampling.

\subsection{Privacy Settings}
We investigated the adoption of restrictive privacy policies by users: our statistical expectation using the Uniform crawler was to acquire $8 \cdot \frac{2^{16}}{2^3} \simeq 65.5K$ users.
Instead, the actual number of collected users was 48.1K.
Because of privacy settings chosen by users, the discrepancy between the expected number of acquired users and the actual number was about 26.6\%.
This means that, roughly speaking, only a quarter of Facebook users adopt privacy policies which prevent other users (except for those in their friendship network) from visiting their friend-list.

\section{Conclusions}
OSNs are among the most intriguing phenomena of the last few years.
In this work we analyzed Facebook, a very popular OSN which gathers more than 500 millions users.
Data relative to users and relationships among users are not publicly accessible, so we resorted to exploit some techniques derived from Web Data Extraction in order to extract a significant sample of users and relations.
As usual, this problem was tackled using concepts typical of the graph theory, namely users were represented by nodes of a graph and relations among users were represented by edges.
Two different sampling techniques, namely BFS and Uniform, were adopted in order to explore the graph of friendships of Facebook, since BFS visiting algorithm is known to introduce a bias in case of an incomplete visit.
Even if incomplete for practical limitations, our samples confirm those results related to degree distribution, diameter, clustering coefficient and eigenvalues distribution.
Future developments involve the implementation of parallel codes in order to speed-up the data extraction process and the evaluation of network metrics.



%
\bibliographystyle{abbrv}
\bibliography{sigproc}  

\begin{thebibliography}{10}

\bibitem{Ahn2007}
Y.~Ahn, S.~Han, H.~Kwak, S.~Moon, and H.~Jeong.
\newblock {Analysis of topological characteristics of huge online social
  networking services}.
\newblock In {\em Proceedings of the 16th international conference on World
  Wide Web}, pages 835--844. ACM, 2007.

\bibitem{albert1999diameter}
R.~Albert.
\newblock {Diameter of the World Wide Web}.
\newblock {\em Nature}, 401(6749):130, 1999.

\bibitem{albert2002statistical}
R.~Albert and A.~Barab{\'a}si.
\newblock {Statistical mechanics of complex networks}.
\newblock {\em Reviews of modern physics}, 74(1):47--97, 2002.

\bibitem{Benevenuto2009}
F.~Benevenuto, T.~Rodrigues, M.~Cha, and V.~Almeida.
\newblock {Characterizing user behavior in online social networks}.
\newblock In {\em Proceedings of the 9th ACM SIGCOMM conference on Internet
  measurement conference}, pages 49--62. ACM, 2009.

\bibitem{Brandes2002}
U.~Brandes, M.~Eiglsperger, I.~Herman, M.~Himsolt, and M.~Marshall.
\newblock {GraphML progress report: Structural layer proposal}.
\newblock In {\em Proc. 9th Intl. Symp. Graph Drawing}, pages 501--512, 2002.

\bibitem{Carrington2005}
P.~Carrington, J.~Scott, and S.~Wasserman.
\newblock {\em {Models and methods in social network analysis}}.
\newblock Cambridge University Press, 2005.

\bibitem{Catanese2010}
S.~Catanese, P.~{De Meo}, E.~Ferrara, and G.~Fiumara.
\newblock {Analyzing the Facebook Friendship Graph}.
\newblock In {\em Proceedings of the 1st Workshop on Mining the Future
  Internet}, pages 14--19, 2010.

\bibitem{Chau2007}
D.~Chau, S.~Pandit, S.~Wang, and C.~Faloutsos.
\newblock {Parallel crawling for online social networks}.
\newblock In {\em Proceedings of the 16th international conference on World
  Wide Web}, pages 1283--1284. ACM, 2007.

\bibitem{Baumgartner2010}
E.~Ferrara, G.~Fiumara, and R.~Baumgartner.
\newblock {Web Data Extraction, Applications and Techniques: A Survey}.
\newblock {\em Tech. Report}, 2010.

\bibitem{Gjoka2010}
M.~Gjoka, M.~Kurant, C.~Butts, and A.~Markopoulou.
\newblock {Walking in facebook: a case study of unbiased sampling of OSNs}.
\newblock In {\em Proceedings of the 29th conference on Information
  communications}, pages 2498--2506. IEEE Press, 2010.

\bibitem{Gjoka2008}
M.~Gjoka, M.~Sirivianos, A.~Markopoulou, and X.~Yang.
\newblock {Poking facebook: characterization of osn applications}.
\newblock In {\em Proceedings of the first workshop on online social networks},
  pages 31--36. ACM, 2008.

\bibitem{Golbeck2006}
J.~Golbeck and J.~Hendler.
\newblock {Inferring binary trust relationships in web-based social networks}.
\newblock {\em ACM Transactions on Internet Technology}, 6(4):497--529, 2006.

\bibitem{Gross2005}
R.~Gross and A.~Acquisti.
\newblock {Information revelation and privacy in online social networks}.
\newblock In {\em Proceedings of the 2005 ACM workshop on Privacy in the
  electronic society}, pages 71--80. ACM, 2005.

\bibitem{Kleinberg2000}
J.~Kleinberg.
\newblock {The small-world phenomenon: an algorithm perspective}.
\newblock In {\em Proceedings of the thirty-second annual ACM symposium on
  Theory of computing}, pages 163--170. ACM, 2000.

\bibitem{Kumar2009}
R.~Kumar.
\newblock {Online Social Networks: Modeling and Mining}.
\newblock In {\em Conf. on Web Search and Data Mining}, page 60558, 2009.

\bibitem{kumar2010structure}
R.~Kumar, J.~Novak, and A.~Tomkins.
\newblock {Structure and evolution of online social networks}.
\newblock {\em Link Mining: Models, Algorithms, and Applications}, pages
  337--357, 2010.

\bibitem{kurant2010bias}
M.~Kurant, A.~Markopoulou, and P.~Thiran.
\newblock On the bias of breadth first search (bfs) and of other graph sampling
  techniques.
\newblock In {\em Proceedings of the 22nd International Teletraffic Congress},
  pages 1--8, 2010.

\bibitem{snap}
J.~Leskovec.
\newblock {Stanford Network Analysis Package (SNAP)}.
\newblock http://snap.stanford.edu/.

\bibitem{Leskovec2008}
J.~Leskovec.
\newblock {\em {Dynamics of large networks}}.
\newblock PhD thesis, Carnegie Mellon University, 2008.

\bibitem{Leskovec2006}
J.~Leskovec and C.~Faloutsos.
\newblock {Sampling from large graphs}.
\newblock In {\em Proceedings of the 12th ACM SIGKDD international conference
  on Knowledge discovery and data mining}, pages 631--636. ACM, 2006.

\bibitem{leskovec2005graphs}
J.~Leskovec, J.~Kleinberg, and C.~Faloutsos.
\newblock {Graphs over time: densification laws, shrinking diameters and
  possible explanations}.
\newblock In {\em Proceedings of the eleventh ACM SIGKDD international
  conference on Knowledge discovery in data mining}, pages 177--187. ACM, 2005.

\bibitem{Liben-nowell2007}
D.~Liben-Nowell and J.~Kleinberg.
\newblock {The link-prediction problem for social networks}.
\newblock {\em Journal of the American Society for Information Science and
  Technology}, 58(7):1019--1031, 2007.

\bibitem{Maia2008}
M.~Maia, J.~Almeida, and V.~Almeida.
\newblock {Identifying user behavior in online social networks}.
\newblock In {\em Proceedings of the 1st workshop on Social network systems},
  pages 1--6. ACM, 2008.

\bibitem{McCown2009}
F.~McCown and M.~Nelson.
\newblock {What happens when facebook is gone?}
\newblock In {\em Proceedings of the 9th ACM/IEEE-CS joint conference on
  Digital libraries}, pages 251--254. ACM, 2009.

\bibitem{milgram1967small}
S.~Milgram.
\newblock {The small world problem}.
\newblock {\em Psychology today}, 2(1):60--67, 1967.

\bibitem{Mislove2007}
A.~Mislove, M.~Marcon, K.~Gummadi, P.~Druschel, and B.~Bhattacharjee.
\newblock {Measurement and analysis of online social networks}.
\newblock In {\em Proceedings of the 7th ACM SIGCOMM conference on Internet
  measurement}, pages 29--42. ACM, 2007.

\bibitem{palmer2002generating}
C.~Palmer and J.~Steffan.
\newblock {Generating network topologies that obey power laws}.
\newblock In {\em Global Telecommunications Conference}, volume~1, pages
  434--438. IEEE, 2002.

\bibitem{Partow}
A.~Partow.
\newblock {General Purpose Hash Function Algorithms}.
\newblock http://www.partow.net/programming/hashfunctions/.

\bibitem{perer2006balancing}
A.~Perer and B.~Shneiderman.
\newblock {Balancing systematic and flexible exploration of social networks}.
\newblock {\em IEEE Transactions on Visualization and Computer Graphics}, pages
  693--700, 2006.

\bibitem{Schneider2009}
F.~Schneider, A.~Feldmann, B.~Krishnamurthy, and W.~Willinger.
\newblock {Understanding online social network usage from a network
  perspective}.
\newblock In {\em Proceedings of the 9th SIGCOMM conference on Internet
  measurement conference}, pages 35--48. ACM, 2009.

\bibitem{Staab2005}
S.~Staab, P.~Domingos, P.~Mike, J.~Golbeck, L.~Ding, T.~Finin, A.~Joshi,
  A.~Nowak, and R.~Vallacher.
\newblock {Social networks applied}.
\newblock {\em IEEE Intelligent systems}, 20(1):80--93, 2005.

\bibitem{Travers1969}
J.~Travers and S.~Milgram.
\newblock {An experimental study of the small world problem}.
\newblock {\em Sociometry}, 32(4):425--443, 1969.

\bibitem{Wilson2009}
C.~Wilson, B.~Boe, A.~Sala, K.~Puttaswamy, and B.~Zhao.
\newblock {User interactions in social networks and their implications}.
\newblock In {\em Proceedings of the 4th ACM European conference on Computer
  systems}, pages 205--218. ACM, 2009.

\bibitem{Ye2010}
S.~Ye, J.~Lang, and F.~Wu.
\newblock {Crawling Online Social Graphs}.
\newblock In {\em Proceedings of the 12th International Asia-Pacific Web
  Conference}, pages 236--242. IEEE, 2010.

\bibitem{Zachary1980}
W.~Zachary.
\newblock {An information flow model for conflict and fission in small groups}.
\newblock {\em Journal of Anthropological Research}, 33(4):452--473, 1977.

\end{thebibliography}
\balancecolumns 

%
%

\end{document}